\documentclass[prl,twocolumn,nofootinbib,reprint]{revtex4-1}
\usepackage{amsmath,amssymb,amsthm,bm,braket,ascmac,bbm}
\usepackage{graphicx}
\usepackage{color}
\usepackage{hyperref}
\usepackage{cancel}
\usepackage{slashed}
\usepackage[table]{xcolor}

\theoremstyle{definition}

\definecolor{mypink1}{rgb}{0.858, 0.188, 0.478}
\definecolor{nicered}{rgb}{.7,.1,.1}

\begin{document}
\hfill KEK-TH-2392

\vspace{0.3cm}

\title{Deeply Learned Preselection of Higgs Dijet Decays at Future Lepton Colliders
\\
}
\author{So Chigusa$^{1,2,3}$, Shu Li$^{4,5,6,7}$, Yuichiro Nakai$^{4,5}$, Wenxing Zhang$^{4,5}$, Yufei Zhang$^{4,5}$ and Jiaming Zheng$^{5,4}$}
\affiliation{\vspace{2mm} \\
$^1$Berkeley Center for Theoretical Physics, Department of Physics,\\
University of California, Berkeley, CA 94720, USA \\
$^2$Theoretical Physics Group, Lawrence Berkeley National Laboratory, Berkeley, CA 94720, USA \\
$^3$KEK Theory Center, IPNS, KEK, Tsukuba, Ibaraki 305-0801, Japan \\
$^4$Tsung-Dao Lee Institute, Shanghai Jiao Tong University, \\
520 Shengrong Road, Shanghai 201210, China \\
$^5$Institute of Nuclear and Particle Physics, School of Physics and Astronomy, Key Laboratory for Particle Physics and Cosmology, Shanghai Jiao Tong University, \\ 800 Dongchuan Road, Shanghai 200240, China \\
$^6$Center for High Energy Physics, Peking University, \\ 5 Yiheyuan Road, Beijing 100871, China \\
$^7$School of Mechanical and Electronic Engineering, Suzhou University, \\ Suzhou 234000, Anhui, China
}

\begin{abstract}
Future electron-positron colliders will play a leading role in the precision measurement of Higgs boson couplings
which is one of the central interests in particle physics.
Aiming at maximizing the performance to measure the Higgs couplings to the bottom, charm and strange quarks,
we develop machine learning methods to improve the selection of events with a Higgs decaying to dijets.
Our methods are based on the Boosted Decision Tree (BDT), Fully-Connected Neural Network (FCNN)
and Convolutional Neural Network (CNN).
We find that the BDT and FCNN algorithms outperform the conventional cut-based method.
With our improved selection of Higgs decaying to dijet events using the FCNN,
the charm quark signal strength is measured with a $16\%$ error, which is roughly a factor of two better than the $34\%$ precision obtained by the cut-based analysis.
Also, the strange quark signal strength is constrained as $\mu_{ss} \lesssim 35$ at the $95\%$ C.L. with the FCNN, which is to be compared with $\mu_{ss} \lesssim 70$ obtained by the cut-based method.
\end{abstract}


\maketitle

{\bf Introduction.--}
The discovery of the Higgs boson at the Large Hadron Collider (LHC)~\cite{ATLAS:2012yve,CMS:2012qbp}
has largely advanced our understanding of the origin of elementary particle masses.
At the same time, the Higgs boson poses major puzzles that must be addressed by physics beyond the Standard Model (SM).
The flavor puzzle asks why the SM fermion masses are hierarchically scattered.
This is just rephrased into the hierarchies of the Yukawa couplings of the fermions
to the Higgs field in the SM.
To address the puzzles and reach a deeper understanding of the origin of masses,
the scrutiny of the nature of the Higgs boson is essential.
The LHC experiment has been searching for Higgs interactions with SM particles.
The results are summarized
in terms of the signal strength
defined as the ratio of the observed production cross section times branching ratio to that predicted by the SM.
To measure or constrain the quark Yukawa interactions,
the LHC has performed various searches such as direct searches for a Higgs boson decaying to a pair of quarks  or with an associated photon.
The current observation/constraint indicates
\cite{ATLAS:2018kot,CMS:2018nsn,CMS:2019hve,ATLAS:2021zwx,ATLAS:2018xfc}
\begin{equation}
\begin{split}
{\rm ATLAS:}&\quad
    \mu_{bb} = 1.02^{+0.12\,+0.14}_{-0.11\,-0.13} \, , \quad
    \mu_{cc} < 26\, ,
    \\[1ex]
{\rm CMS:}&\quad
    \mu_{bb} = 1.01\pm 0.22 \, , \qquad \,\,
    \mu_{cc} < 70\,,
\end{split}
\end{equation}
for the decay modes $H \rightarrow b \bar{b}$ and $H \rightarrow c \bar{c}$, respectively.
Regarding the Yukawa interactions of lighter quarks~\cite{Gao:2016jcm}, the direct observation of a Higgs boson decay to such quarks
is extremely challenging at the LHC.
In fact, the strange quark Yukawa coupling is constrained by the exclusive $H\to \phi \gamma$ decay
\cite{Kagan_2015, K_nig_2015, Perez_2016},
which sets the limit
\cite{ATLAS:2017gko}
\begin{align}
    \mu_{ss} \lesssim 7.2\times 10^8 \, ,
\end{align}
assuming the SM expectation for the Higgs-photon interaction.
Future electron-positron colliders such as
the Circular Electron Positron Collider (CEPC)~\cite{CEPCStudyGroup:2018ghi},
Future Circular Collider in electron mode (FCC-ee)~\cite{Proceedings:2019vxr}
and International Linear Collider (ILC)~\cite{ILC:2007oiw}
have advantages over the LHC in terms of
the low background and well-defined initial state
and are expected to significantly improve the precision measurement of Higgs boson interactions.
Then, it is of critical importance to maximize the search power of future colliders
to accelerate the discovery of Higgs decay modes inaccessible at the LHC
and to find any small deviation from the SM prediction that gives direct evidence of new physics.

There are two major steps in the measurement of the quark Yukawa coupling via direct searches for a Higgs boson decaying
into a pair of quarks:
the preselection to separate Higgs decaying to dijet ($H \rightarrow jj$) events
from all non-$H \rightarrow jj$ background events
and the subsequent jet flavor tagging.
Regarding jet flavor tagging,
bottom and charm quark jets are tagged by identifying displaced charged track vertices
\cite{ATLAS:2015thz,CMS:2017wtu}.
The strange jet tagging has been discussed in refs.~\cite{Nakai:2020kuu,Erdmann:2020ovh} for the LHC and
in ref.~\cite{Duarte-Campderros:2018ouv} for future $e^+e^-$ colliders.
In addition, recently, applications of machine learning (ML) techniques, in particular deep learning,
to jet flavor tagging have been under intense investigation
(see ref.~\cite{Larkoski:2017jix} and references therein),
and the improvement of the tagging performance has been reported.
For instance, the Convolutional Neural Network (CNN) has been used to analyze jet substructures
and identify boosted Higgs production modes;
the production of $t\bar{t}h$ has been studied with extreme gradient boosted trees
and neural network models~\cite{Chung:2020ysf, Lin:2018cin, Santos:2016kno}.
On the other hand, the preselection of $H \rightarrow jj$ events at future $e^+e^-$ colliders
was studied by using the conventional cut-based method
\cite{Ono:2013sea}.
In the work presented in the talk~\cite{Bai:2016},
the Boosted Decision Tree (BDT) was used for the preselection only after the cut-based selection.
Considering the success of ML applications in the jet flavor tagging,
it is natural to test various ML algorithms including deep learning
to improve the preselection,
aiming at maximizing the performance to measure the Higgs couplings to the bottom, charm and strange quarks.

In the present letter, we develop ML methods based on the BDT, Fully-Connected Neural Network (FCNN)
and Convolutional Neural Network (CNN)
with the aim to improve the preselection of $H \rightarrow jj$ events at future $e^+e^-$ colliders.
We find that the BDT and FCNN-based algorithms outperform the cut-based method
and that combined with the BDT introduced in ref.~\cite{Bai:2016}.
The resulting sensitivity to the signal strength for each Higgs decay mode of $H \rightarrow b \bar{b}$,
$H \rightarrow c \bar{c}$ or $H \rightarrow s \bar{s}$ is then studied.

\begin{table*}[!t]\centering
	\begin{tabular}{|c||c|c|c|c|c|c|c|} \hline
		& \multicolumn{2}{c|}{Signal} & \multicolumn{5}{c|}{Background}     \\ \hline
		Process & $HZ(\to\nu\bar{\nu})$ & $\nu \bar{\nu} H \, (WW \, \text{fusion}) $ & $ZZ$ & $W^+W^-$ & $q\bar{q}$ & $e^\pm \nu W^\mp$ & $e^+ e^- Z$ \\ \hline
		Cross section in pb & 
		0.0469 & 0.00774 & 1.03 & 15.4 & 50.2 & 5.14 & 4.73          \\ \hline
		Number of events  & 11725 & 1942 & 257250 & 3.85$\times 10^6$ & 1.255$\times 10^7$ & 1.285$\times 10^6$ & 1.182$\times 10^6$          \\ \hline
	\end{tabular}
	\caption{\label{tab::cs} The cross section and number of events
		of each signal/background process for the $e^+ e^-$ collision with $\sqrt{s}=250\,\mathrm{GeV}$
		and an integrated luminosity of $\mathcal{L}=250\,\mathrm{fb}^{-1}$.}
\end{table*}

{\bf Event generation.--}\label{sec::cut_based}
In order to investigate the preselection performance,
we assume an $e^+e^-$ collider with unpolarized $e^{+}e^{-}$ beams, the center-of-mass energy $\sqrt{s}=250\,\mathrm{GeV}$
and the integrated luminosity $\mathcal{L}=250\,\mathrm{fb}^{-1}$.
Our focus is on the signal process with the $\nu \bar{\nu}H
(\to jj)$ final state,
either through the Higgs-strahlung $e^{+} e^{-} \to Z H$ with $Z \to \nu \bar{\nu}$
or the $WW$ fusion, which possesses the highest sensitivity to $H \to j j$
\cite{CEPCStudyGroup:2018ghi, Ono:2013sea, Bai:2016, Abramowicz_2017}.
The signal process is characterized by two hard jets and a large missing transverse energy (MET)
corresponding to the sum of two neutrino transverse momenta.
We include all Higgs decay channels when generating the Higgs events.
The preselection methods described below will mostly pick up $H\rightarrow jj$ events.
The main background processes are $e^+ e^- \to Z Z$ and $W^+ W^-$,
both of which result in $\rm jets + MET$ when one of the vector bosons
decays leptonically and the other decays hadronically.
The remaining background processes
are $e^+ e^- \to q \bar{q}$, $e^\pm \nu W^\mp$ and $e^+ e^- Z$, whose cross sections are not negligible.
The cross section and the corresponding number of events of each signal/background process
are listed in Table~\ref{tab::cs}.

Our simulation uses MadGraph5\_aMC@NLO
 v2.6~\cite{Alwall:2014hca}
for Monte Carlo sample generation of hard scattering processes,
PYTHIA8.2~\cite{Sjostrand:2006za, Sjostrand:2014zea} for parton showering and hadronization
and DELPHES v3.4~\cite{deFavereau:2013fsa} with FastJet v1.0~\cite{Cacciari:2011ma} for jet clustering and detector simulation,
where the default card of the CEPC in DELPHES v3.4 and default settings in PYTHIA8 are used.

\begin{table*}[!t]\centering
	\begin{tabular}{|c||c|c|c|c|c|c|c|} \hline
	& \multicolumn{2}{c|}{Signal} & \multicolumn{5}{c|}{Background}     \\ \hline
	    Process & $HZ(\to\nu\bar{\nu})$ & $\nu \bar{\nu} H \, (WW \, \text{fusion}) $ & $ZZ$ & $W^+W^-$ & $q\bar{q}$ & $e^\pm \nu W^\mp$ & $e^+ e^- Z$ \\ \hline
		 Before cut  & 11725 & 1942 & 275250 & 3.85$\times 10^6$ & 1.255$\times 10^7$ & 1.285 $\times 10^6$ & 1.1825 $\times 10^6$
  \\ \hline
  $80\,\mathrm{GeV}< M_{\rm miss} < 140\,\mathrm{GeV}$ & 8854 &1322 & 83565 &409174 & 33876 &242224 &241020  \\ \hline
  $20\,\mathrm{GeV}< P_T < 70\,\mathrm{GeV}$ & 8161 &1072 &49099 &291164 & 4376 &169402 &144559  \\ \hline
  $|P_L| < 60\,\mathrm{GeV}$ & 7967 &969 &16086 &145018 & 4043 &83310 &38178  \\ \hline
  $N_{\rm chd}\geq 10$  & 7772 &946 &14072 &53070 & 4009 &4478 &0  \\ \hline
  $P_{\rm max}< 30\,\mathrm{GeV}$ & 6963 &855 &10951 &27265 & 2619 &447 &0  \\ \hline
  $Y_{23}< 0.02$ & 4623 &554 &7546 &4344 & 2193 &109 &0  \\ \hline
  $0.2 < Y_{12} < 0.8$ & 4535 &500 &4995 &3385 & 2008 &91 &0  \\ \hline
  $100\,\mathrm{GeV}< M_{jj} < 130\,\mathrm{GeV}$ & 4331 &475 &856 &1677 & 277  &50 &0  \\ \hline
 \end{tabular}
	\caption{\label{tab:cutflow} The cut-flow of the preselection using the variables presented in the main text.
    }
\end{table*}

{\bf Cut-based method.--}
We perform a cut-based preselection of $H \rightarrow jj$ events to set a benchmark for comparison.
We adopt the same cut variables and criteria as those used in ref.~\cite{Ono:2013sea}, which are summarized as follows:
\begin{itemize}
	\item Missing mass $M_{\rm miss}$

	Since $e^+ e^- \to Z H$ is the dominant signal process, we consider a cut
	that picks up most of $ZH$ events while reducing background events.
	We then use the missing mass defined as
	\begin{equation}
    	M_{\rm miss} \equiv \sqrt{\left(\sqrt{s}-E_{\rm vis} \right)^2-\slashed{P}_T^2} \, ,
	\end{equation}
	where $E_{\rm vis}$ is the total energy of visible particles in an event
	and $\slashed{P}_T$ is the total missing transverse momentum.
	For the $Z H$ channel, $\slashed{P}_T$ is ideally equal to the transverse momentum of the Higgs boson
	and $M_{\rm miss}$ peaks at the $Z$ boson mass.
	Thus, we require $80\,\mathrm{GeV}\leq M_{\rm miss} \leq 140\,\mathrm{GeV}$ as our selection criteria.

	\item $P_{T/L}$, $P_{\rm max}$, $N_{\rm chd}$

	To reduce the $q\bar{q}$ background, we utilize the total transverse/longitudinal momentum of visible particles $P_{T/L}$
	and the maximum track momentum $P_{\rm max}$ in an event.
	We require $20\,\mathrm{GeV}< P_T < 70\,\mathrm{GeV}$, $|P_L| < 60\,\mathrm{GeV}$ and $P_{\rm max}< 30\,\mathrm{GeV}$
	as our criteria.
	On the other hand, to well reduce the leptonic ($\ell \ell \ell \ell$) background,
	we set a cut on the number of charged tracks, $N_{\rm chd}> 10$.

	\item $Y_{12}, Y_{23}$

	We apply the $k_T$ algorithm for the jet clustering,\footnote{
	\color{black}
	we use the variables $Y_{12}$ and $Y_{23}$ to extract events with exactly two hard partons, which correspond to two quarks from the Higgs decay for the signal processes.
    To correctly count the number of hard partons, it is necessary to use the $k_T$ clustering algorithm.
	}
	which defines the following variables 
	\cite{Catani:1991hj,Catani:1993hr,Ellis:1993tq, CMS-PAS-JME-09-001}:
	\begin{equation}
	y_{ij}=\frac{2\min(E_i^2, E_j^2)(1-\cos \theta_{ij})}{E_{\rm vis}^2} \, ,
	\end{equation}
	where $E_{i/j}$ denotes the energy of a (pseudo)particle labeled by $i/j$
	and $\theta_{ij}$ is the angle between the momenta of $i$ and $j$.
	Two (pseudo)particles are clustered when the $y$-value satisfies $y_{ij} < y_{\rm cut}$.
	Then, $Y_{12}$ and $Y_{23}$ are defined as the maximum and minimum values of $y_{\rm cut}$ with which an event contains exactly two jet in the final state, respectively.
	We require $Y_{23}< 0.02$ and $0.2 < Y_{12} < 0.8$ to further reduce background events mainly from the $W^{+}W^{-}$ channel.

	\item Di-jet mass $M_{jj}$

	We use the invariant mass of the two hardest jets in the final state $M_{jj}$,
	which is ideally equal to the Higgs mass for the signal events.
	Our requirement is $100\,\mathrm{GeV}< M_{jj} < 130\,\mathrm{GeV}$.

\end{itemize}
All of these variables are also used as inputs for the BDT and FCNN presented below.

Table~\ref{tab:cutflow} shows the cut-flow.
Compared with ref.~\cite{Ono:2013sea}, we consider more general background processes including $e^+ e^- \to e^\pm \nu W^\mp$ and $e^\pm e^\mp Z$.
The main background events remaining after the cuts are $e^+ e^- \to W^+ W^-$ and $ZZ$.
They could be further reduced by a harder cut on $Y_{ij}$.
However, it also rejects a non-negligible fraction of signal events, which degrades the performance of the preselection.

{\bf BDT.--}
In order to take account of correlations among the variables listed in the cut-based method,
we utilize a BDT algorithm whose inputs are those used in the cut-based method.
The BDT is implemented with the {\tt scikit-learn} library~\cite{DBLP:journals/corr/abs-1201-0490},
and its outputs are used as the unique variables for signal/background classification,
or as a variable with which we further reduce background events on top of the cut-based preselection.
The decision tree classifier has a maximum depth of $25$ and requires a minimum of $30$ samples for a leaf node. It is fitted by the AdaBoost-SAMME\,\cite{Adaboost} algorithm with a learning rate of $0.1$ and the maximum number of estimators is set to 75. These hyperparameters are manually chosen towards the best Higgs signal significance.

\begin{table*}[htbp]\centering
 \begin{tabular}{|c||c|c|c|c|c|c|c||c|} \hline
  & \multicolumn{2}{c|}{Signal} & \multicolumn{5}{c||}{Background} &  Significance  \\ \hline
  Process & $HZ(\to \nu\bar{\nu})$ & $\nu \bar{\nu} H \, (WW \, \text{fusion}) $ &  $W^+ W^-$ & ~~$q\bar{q}$~~ &  ~~$ZZ$~~ & $e^\pm \nu W^\mp$ & $e^+ e^- Z$ &  \\ \hline
  Cut-based & 4331 &475 &1677  &277 &856 &50 &0 &54.9$\, \sigma$ \\ \hline
  BDT-only &6721 & 1047&61& 195 & 399 & 9 & 0 &84.6$\, \sigma$\\ \hline
  FCNN-only &9562  & 1427&754 & 1101 &419 & 109 & 1 &95.0$\, \sigma$ \\ \hline
  CNN-only & {4776} & {749} & {6337} & {5407} & {432} & {1237} & {552}  & {39.6$\, \sigma$}\\ \hline
  Cut+BDT & 3744 &403 & 10  &129 &57 &1 &0 &62.9$\, \sigma$ \\ \hline
  Cut+FCNN & 3941 &409 &95  &145 &170 &0 &0 & 63.1$\, \sigma$ \\ \hline
  {Cut+CNN}& {3733} &{409} &{289}  &{219} &{486} &{0} &{0} &{57.8$\, \sigma$} \\ \hline
 \end{tabular}
 \caption{\label{tab:cut_ML} The number of signal/background events and signal significance
 after each preselection method.
 BDT-only denotes the method that the output of the BDT is used as a unique variable to classify signal/background events.
 The same applies to FCNN-only and CNN-only.
 In Cut+BDT, the output of the BDT is used as a cut variable to reduce background events on top of the cut-based preselection.
 The same applies to Cut+FCNN and Cut+CNN.
 }
\end{table*}

{\bf FCNN.--}
We implement the FCNN algorithm by using the {\tt Keras 2.7.0}~\cite{DBLP:journals/corr/MerrienboerBDSW15} application programming interface\,(API) with the {\tt Tensorflow 2.7.0}~\cite{abadi2016tensorflow} backend.  
We utilize a simple neural network in which every neuron in one layer connects to all neurons in the subsequent layer.
In principle, this fully-connected neural network with a sufficient number of hidden layers
has the ability to approximate any continuous function in a finite-dimensional space~\cite{Hornik1988} and has a potential to learn features indistinguishable by the BDT.
In our study, the variables used in the cut-based method are fed into the FCNN.
We have tested various architectures and
found that the best performance is achieved by a network with 9 hidden layers
where the first 2 layers contain 100 neurons and the rest contain 90 neurons. We note that the performance of this network is only improved marginally beyond a smaller FCNN with 4 layers.
We employ the Rectified Linear Unit (ReLU) as the activation function for the hidden layers
and the softmax function for the output layer with two neurons.
The optimizer is taken to be ADAM with the {\tt Keras} default parameters  and the loss function is cross entropy.
The network outputs a score of the likelihood for an input event recognized as a $H \rightarrow jj$ signal event,
which is used as the unique variable to classify signal/background events, or as a cut variable to reduce background events on top of the cut-based preselection as in the case of the BDT.

{\bf CNN.--}
In addition to the BDT and FCNN, we test the CNN algorithm whose inputs are 2D event images.
The image of an event has 3 ``colors"~\cite{Komiske:2016rsd} : particle momenta, particle charges and jet momenta.
All the momenta are normalized by multiplying $1/\sqrt{s}$ for the stability of the network.
2D images are expanded in terms of the $\eta \times \phi$ coordinate system.
The number of pixels for each channel is 33$\times$36
where the size of a pixel is given by $\Delta \eta = 0.1 \, (0.4)$ for $|\eta| < 1.0 \, (1.0 < |\eta| < 3.4)$
and $\Delta \phi = 2\pi /36$.
For the particle (jet) momentum color, a pixel value is defined as
the vectorial summation of particle (jet) momenta inside the pixel, normalized by the collision energy.
For the jet momentum color, the jet clustering is conducted by the anti-$k_T$ algorithm~\cite{Cacciari:2008gp}.
We have also tested the $b$-tagging label as the fourth color but the performance is similar for 3 and 4 colors.
The $b$-tagging pixel value is defined as the number of tight $b$-jets recognised by the anti-$k_T$ algorithm.
The CNN is implemented with the {\tt TensorFlow}\,\cite{abadi2016tensorflow} framework and the {\tt Keras}\,\cite{DBLP:journals/corr/MerrienboerBDSW15} API. 
The network architecture we utilize is summarized as follows.
An input image with 3 or 4 colors for an event enters three convolutional layers each of which has 200 filters
with $4\times 4$ kernel.
A max-pooling layer with $2 \times 2$ reduction follows.
The generated feature maps are coupled to the fourth convolutional layer of 200 filters with $4\times 4$ kernel.
The 2D image is then flattened and coupled through fully-connected layers with 100, 100, 80, 80, 60 and 60 neurons respectively.
The activation function is the ReLU for the hidden layers
and the softmax function for the output layer with two neurons.
The optimizer is Adadelta and the loss function is cross entropy.
The optimizer is ADADELTA\,\cite{ADADELTA} with an initial learning rate of $0.3$ and
other parameters are set as the {\tt Keras} default.
As in the case of the BDT and FCNN, the network output is used as the unique variable to classify signal/background events,
or as a cut variable on top of the cut-based selection.

\begin{figure}[!t]\centering
	\includegraphics[width=0.48\textwidth]{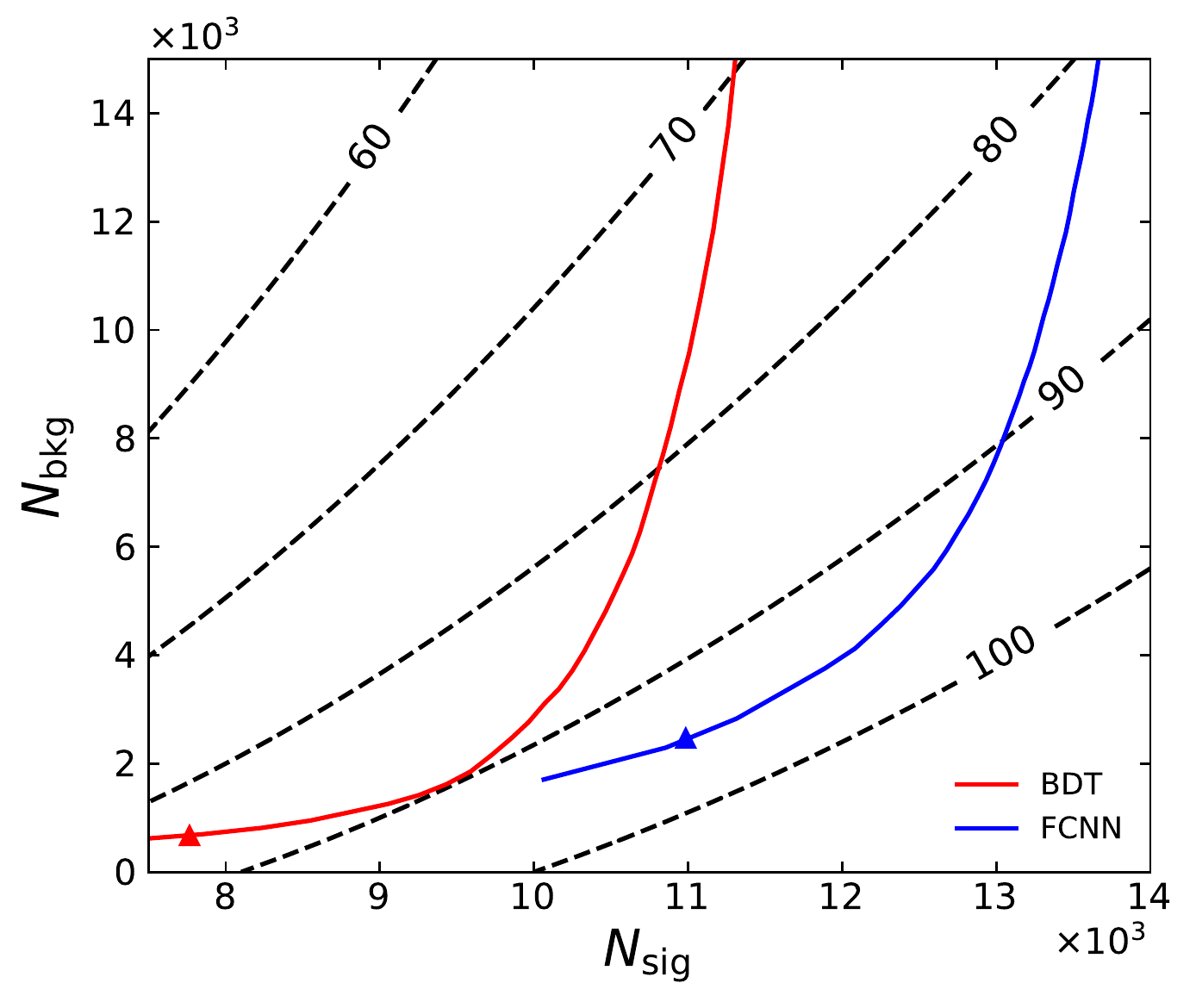}
	\caption{\label{fig:sigeff}
	The number of signal ($N_{\rm sig}$) and background ($N_{\rm bkg}$) events
	that passes the preselection using the BDT (red solid) and FCNN (blue solid).
	The triangle marker on the curve represents the number of events obtained by the classification threshold utilized
	in Table~\ref{tab:cut_ML}.
	The black dashed curves denote contours of the signal significance.
	}
\end{figure}

\begin{table*}[!t]\centering
	\setlength{\tabcolsep}{3mm}
	\begin{tabular}{|c||c|c|c|c|c|} \hline
		&Cut-based & BDT-only & FCNN-only & Cut+BDT & Cut+FCNN
		\\ \hline
		$\mu_{bb}$ & $1\pm 0.021$  & $1\pm 0.016$  & $1\pm 0.013$ &  $1\pm 0.021$ & $1\pm 0.021$  \\ \hline
		$\mu_{cc}$ & $1\pm 0.34$ & $1\pm 0.16$  & $1\pm 0.16$ &  $1\pm 0.21$ & $1\pm 0.22$  \\ \hline
		$\mu_{ss}$ & $70$    & $36$    & $35$   &  $47  $ & $48  $  \\ \hline
	\end{tabular}
	\caption{\label{tab:yukawa_signal} The expected $1\sigma$ errors of the signal strength for the bottom ($\mu_{bb}$)
	and charm ($\mu_{cc}$)
	Yukawa couplings and the $95\%$ C.L. upper limits on that of the strange quark ($\mu_{ss}$)
	with various preselection methods presented in Table~\ref{tab:cut_ML}, assuming statistical uncertainty only.}
\end{table*}

{\bf Results.--}
Table~\ref{tab:cut_ML} shows the number of signal ($N_{\rm sig}$) and background ($N_{\rm bkg}$)
events remaining after each preselection method.
We also estimate the signal significance $N_{\rm sig}/\sqrt{N_{\rm bkg}+N_{\rm sig}}$
for comparison.
The classification threshold of the BDT/FCNN/CNN output is chosen to obtain a large signal significance.
The most suitable choice of the threshold depends on a physics variable that we would like to extract,
e.g., the Yukawa coupling,
because sensitivities to different variables are generally given by different combinations of $N_{\rm sig}$ and $N_{\rm bkg}$.

We can see from Table~\ref{tab:cut_ML}
that the BDT-only method outperforms the cut-based method in terms of the signal significance,
with more signal events and less background events after the preselection in each channel.
We also find that the significance of the Cut+BDT method is worse than that of the BDT-only
because the cut conditions are so strict that the number of signal events is excessively reduced.
In fact, Table~\ref{tab:cutflow}
shows that the significance could be no longer larger than $70\,\sigma$ after the cut-based preselection
even if the BDT distinguished all background events from signal events.
We note, however, that the presented signal significance does not faithfully reflect
the performance of a specific application of the Higgs preselection such as the Yukawa coupling measurement.

The FCNN-only method shows the best signal significance among all preselection methods in Table~\ref{tab:cut_ML},
which results from the highest signal acceptance and a high background rejection rate.
The Cut+FCNN method is less efficient than the FCNN-only due to the similar reason discussed for the case of the BDT.

Figure~\ref{fig:sigeff} demonstrates the dependence of the BDT and FCNN results on the choice of the threshold.
The relationship between $N_{\rm sig}$ and $N_{\rm bkg}$ for the BDT and FCNN analyses
is plotted by scanning the threshold value.
We can see that $N_{\rm sig}$ of the FCNN is larger than that of the BDT for any fixed value of $N_{\rm bkg}$,
indicating a clear advantage of the FCNN.

Most events that pass the preselection by the BDT or FCNN, regardless of being a true Higgs event or not,
have di-jet masses within $120\sim125.2$\,GeV.\footnote{\color{black}
With the neural networks discussed in this work, the background is solely estimated from their performance on the simulation data. A pratical bump search usually preforms a sideband fit for better control of background estimation and systematic uncertainties. The combination of the NNs and the sideband subtraction method will require additional treatment\cite{Englert:2018cfo,Wunsch:2019qbo,Bradshaw:2019ipy,Kasieczka:2020yyl,Benkendorfer:2020gek,Kitouni:2020xgb,Ghosh:2021hrh} to decorrelate the NN from resonance-sensitive variables such as the invariant mass.}
Furthermore, since the flavor information is not used in training,
all quark flavors are equally likely to pass the preselection.
Indeed, we have fed equal numbers of events from $h\rightarrow b\bar b$, $h\rightarrow c\bar c$
and $h\rightarrow g g$ into the trained BDT and FCNN and observed almost equal passing rates.
Therefore, the tested BDT and FCNN architectures mostly select $H\rightarrow q\bar{q}$ signals in a flavor-blind manner,
which is crucial for measuring the quark Yukawa couplings.

The signal significance of the CNN-only method is observed to be much worse than the other methods
while the Cut+CNN is only slightly better than the cut-based method.
The situation is the same for all the other CNN architectures we have tested. The CNN is trained with more than $10^5$ sample events that contain equal numbers of signal and background events.
We monitor the training of the CNN by its accuracy on the training and validation sets with a classification threshold of $0.5$. We have observed that the classification accuracy plateaus near $90\%$ during the training,
which signals the saturation of the CNN performance.
However, increasing the classification threshold for the tested CNN does not improve the Higgs signal significance as sharply as that for the BDT or the FCNN,
since the tested CNN assigns high scores for some background events just like for the signal events.
As a result, the CNN gives the worst performance among the machine learning methods, \textcolor{black}{and its training accuracy always collapses after going through several epochs even if we use smaller amount of training data.}
It is hard to decipher the CNN black box, and we only provide an educated guess for such an under-performance of the tested CNN as follows.
Although the CNN is renowned for its ability to recognize shapes and edges in an image,
it is not clear whether the CNN is equally good at perceiving precise numerical correlations over large separation in an image.
The latter type of information such as the di-jet mass and missing mass are used directly
as input variables of the BDT and FCNN.
However, it may be hard for the CNN to recognize these variables because it takes several convolutions to correlate two far-separated points
while the numerical precision might be lost during pooling.
We thus conclude that the CNN architecture presented above is not a suitable ML technique for the Higgs preselection
and leave a possible improvement of the performance to a future study.

{\bf Yukawa coupling measurements.--}
To showcase the importance of the improved preselection by the BDT and FCNN,
we estimate the resultant improvement on sensitivities to the bottom, charm and strange Yukawa couplings
measured or constrained by $H \rightarrow f\bar{f}$ ($f$ is a quark) searches in the absence of systematic uncertainty.
Assuming the SM value for the Higgs production cross section,
the signal strength is given by
\begin{equation}
\mu_{ff} = \frac{{\rm Br} \left(H\rightarrow f\bar f \right) }
{{\rm Br}^{\rm SM} \left(H\rightarrow f\bar f \right) }\, .
\end{equation}
Here, ${\rm Br}^{\rm SM} \left(H\rightarrow f\bar f \right)$
denotes the SM expectation of the branching ratio for $H \rightarrow f\bar{f}$.
We now define the number of events that pass the preselection of $H \rightarrow jj$ events
and the subsequent quark flavor tagging
as $N_f=S_f+B_f$
where $S_f \, (B_f)$ is the number of signal (background) events.
They are related to their SM expectation values via $S_f\simeq \mu_{ff} S_f^{\rm SM}$ and $B_f\simeq B_f^{\rm SM}$.
The statistical significance of deviation from the SM is characterized by
\begin{equation}
\chi^{2}\equiv\frac{\left(N_f-N_f^{{\rm SM}}\right)^{2}}{N_f^{{\rm SM}}}
\simeq
\left(
\frac{(\mu_{ff}-1)S_f^{\rm SM}}{\sqrt{S_f^{\rm SM}+B_f^{\rm SM}}}
\right)^2
\,,\label{eq:sig_def}
\end{equation}
where $\chi\simeq 1 \, (1.96)$ corresponds to the $1\sigma$ ($95\%$\,C.L.) deviation.
Since the bottom and charm Yukawa couplings are expected to be measured at future lepton colliders,
we assume that both the true and observed central values are $\mu_{ff}=1 \, (f=b,c)$ and estimate the $1\sigma$ errors by requiring $\chi<1$.
On the other hand, for the strange Yukawa coupling, we obtain the $95\,\%$ C.L. upper bound on $\mu_{ss}$ by requring $\chi < 1.96.$

For the bottom and charm quark tagging, we assume a set of tagging efficiencies presented in a CEPC study\,\cite{Ruan:2018yrh}
where an inclusive $Z\rightarrow q\bar q $ sample is used.\footnote{
		For simplicity, we assume that the tagging efficiencies for $H\to q\bar{q}$ events are comparable to those of ref.~\cite{Ruan:2018yrh}.
		The purpose of our analysis is to compare the performance of various machine learning approaches to the Yukawa coupling measurements, and a more rigorous treatment of the flavor tagging is beyond the scope of the current paper.
}
The $b$-jet tagging efficiency, mistagging rate of the $c$-jet and mistagging rate of the other light quark or gluon jet are
$\epsilon_b=80\%$, $\epsilon_c^{\rm bkg}=8\%$ and $\epsilon_q^{\rm bkg}=1\%$, respectively.
Similarly, the $c$-jet tagging efficiency, mistagging rate of the $b$-jet
and mistagging rate of the other light quark or gluon jet are
$\epsilon_c=60\%$, $\epsilon_b^{\rm bkg}=14\%$ and $\epsilon_q^{\rm bkg}=7\%$, respectively.
The strange jet tagging is much less efficient.
For a rough estimate, we adopt the tagging method presented in ref.~\cite{Duarte-Campderros:2018ouv}
where the efficiency of tagging two signal $s$-jets, mis-tagging rate of two background jets from a $H \rightarrow jj$ process
and mis-tagging rate of two jets from a non-$H \rightarrow jj$ process are
$\epsilon_{ss}=96\%$, $\epsilon_{\rm bkg}^{H\rightarrow jj}=26\%$ and $\epsilon_{\rm bkg}^{{\rm non-}jj}=70\%$, respectively.

In table~\ref{tab:yukawa_signal}, we summarize the expected $1\sigma$ errors of the signal strength
for the bottom and charm quarks and the $95\%$\,C.L. upper limits on that of the strange quark
with various preselection methods according to the performances listed in Table~\ref{tab:cut_ML}, assuming only statistical uncertainty.
For a conservative estimate, we have assumed that a background event is counted in $B_f^{\rm SM}$
whenever it contains two jets that are tagged as the flavor of our concern.
The table indicates that the cut-based method reaches a good precision of $2\%$ for the bottom Yukawa,
while the BDT and FCNN improve it marginally.
The BDT and FCNN preselection methods give a similar measurement precision of $\mu_{cc}$, roughly $16\%$,
which improves significantly from the cut-based result of $34\%$.
The precision is weakened in the combined analysis of the Cut+BDT or Cut+FCNN,
which determines the signal strength $\mu_{cc}$ with a $21\%$ or $22\%$ error, respectively.
We can see that the power of ML methods in rejecting backgrounds is essential for the charm-Yukawa precision measurement.
For the strange Yukawa, the BDT and FCNN preselection methods result in $95\%$ C.L. upper bounds, $\mu_{ss}\lesssim 36$
and $\mu_{ss}\lesssim 35$ respectively,
which are again significantly better than the cut-based result of $\mu_{ss}\lesssim 70$.

We have shown that the ML methods can significantly reduce the statistical error in the charm and strange Yukawa measurements. On the other hand, the ignored systematic uncertainty, including those from the tagging efficiencies, event simulations and signal/background modeling have to be included in the limit setting.
A further treatment of systematic uncertainty is important for the bottom and charm Yukawa measurements
where statistical uncertainty is small, as shown in Table~\ref{tab:yukawa_signal}.
For the strange Yukawa, the dominant uncertainty may still be statistical.  


{\bf Conclusions.--}
We have studied how ML can help improve the performance for
the preselection of $H \rightarrow jj$ events at future $e^+e^-$ colliders.
The BDT, FCNN and CNN were applied to select $H \rightarrow jj$ events
from a large amount of backgrounds of $ZZ$, $W^+W^-$, $q \bar{q}$, $e^\pm \nu W^\mp$ and $e^+ e^- Z$.
Our result indicates that the FCNN performs the best in improving the Higgs signal significance
and is significantly better than the cut-based method,
while the BDT has a slightly weaker but comparable capability.
On the contrary, the CNN we have used is not suitable for the Higgs preselection task, \textcolor{black}{and the result of CNN is suboptimal. The performance is not improved even if we take less or more training data.}
We also tested the performance of combinations of the ML methods with the cut-based selection.
Such combinations degrade the resulting signal significance of the BDT and FCNN since only a limited number of signal events pass the cut-based selection.
Nevertheless, the combined methods may be useful if one needs strict control of probable background events.

The precision measurement of the quark Yukawa couplings
is one of the most important missions at future lepton collider experiments.
The Higgs preselection based on the BDT and FCNN algorithms improves sensitivities to the charm and strange Yukawa couplings significantly
compared to the conventional cut-based analysis,
while only a mild improvement has been observed for the bottom Yukawa coupling.
In summary, the BDT and FCNN algorithms are highly useful to select $H \rightarrow jj$ events
and improve direct measurements on the light quark Yukawa couplings at future lepton colliders.

%
\section*{Acknowledgements}

We would like to thank Gang Li, Kun Liu, Hiroaki Ono, Manqi Ruan, Junfeng Wu and Dan Yu for the discussions.
SC was supported by JSPS KAKENHI Grant No. 20J00046.
SC was supported by the Director, Office of Science,
Office of High Energy Physics of the U.S. Department of Energy under the Contract No.~DE-AC02-05CH1123.
YN is supported by Natural Science Foundation of China under grant No.~12150610465.

\vspace{0.5cm}

\phantomsection
\addcontentsline{toc}{section}{References}
\bibliography{higgsRef}
\bibliographystyle{utphys}

\end{document}